# Towards systems tissue engineering: elucidating the dynamics, spatial coordination, and individual cells driving emergent behaviors

Matthew S Hall[1], Joseph T Decker[1], Lonnie D Shea[1*]

[1]Department of Biomedical Engineering, University of Michigan, Ann Arbor MI
[*]Address correspondence to Lonnie D. Shea (ldshea@umich.edu)

## ABSTRACT

Biomaterial systems have allowed for the *in vitro* production of complex, emergent tissue behaviors that were not possible with conventional 2D culture systems allowing for analysis of the normal development as well as disease processes. We propose that the path towards developing the design parameters for biomaterial systems lies with identifying the molecular drivers of emergent behavior through leveraging technological advances in systems biology, including single cell omics, genetic engineering, and high content imaging. This research focus, which we term systems tissue engineering, can uniquely interrogate the mechanisms by which complex tissue behaviors emerge with the potential to capture the contribution of i) dynamic regulation of tissue development and dysregulation, ii) single cell heterogeneity and the function of rare cell types, and iii) the spatial distribution and structure of individual cells and cell types within a tissue. Collectively, systems tissue engineering can facilitate the identification of biomaterial design parameters that will accelerate basic science discovery and translation.

**MAIN TEXT**

**1. Introduction**

Tissues are composed of multiple cell types that reside within a complex, continuously changing three-dimensional microenvironment consisting of numerous inputs, including extracellular matrix (ECM), soluble factors, mechanical forces, and cell-cell contacts, all of which combine to drive collective tissue function (1, 2). By mimicking and reproducing these inputs, biomaterial and microsystems technologies (3, 4) have allowed for the *in vitro* production of diverse emergent tissue behaviors that would not otherwise be possible with conventional 2D culture systems, such as vasculogenic microcapillary networks (5-7), beating cardiomyocyte microtissues (8, 9), functional skeletal muscle (10, 11), and stem cell-derived tissue organoids (12, 13). The tunable nature of biomaterial microsystems also allows for engineering the microenvironment to interrogate the contribution of specific properties on the resulting cell and tissue behavior. These in vitro biomaterial microsystems therefore have incredible potential for molecularly dissecting tissue formation, the disease initiation and progression, or the mechanism of action for therapeutic compounds (3, 14). Leveraging the potential of these complex platforms requires the ability to identify the design parameters that direct the cell processes and tissue formation.

The path towards developing improved microsystems and material platforms will likely involve the ability to integrate systems biology with the analysis of tissue dynamics and structure. Historically, the identification of biomaterial design parameters has ranged from reductionist approaches to high throughput screening systems, which have frequently involved a focused number of cell types or cellular responses, failing to capture the complex multivariate biology present in vivo (15). More recent approaches have sought to capture increased complexity through co-culture to multi-culture systems, 2D and/or 3D culture across multi-well plates, and microphysiological systems that integrate multiple organ systems. Native *in vivo* tissues are increasingly analyzed with emerging molecular tools that provide measurements of thousands/millions of factors in a single experiment. These tools, including single cell RNA sequencing, have revealed the vast heterogeneity in molecular phenotype between cells of the same classical type even within the same tissue compartment (16, 17). The large data sets generated have the potential to identify the interconnected responses inherent to spatially coordinated biological systems, the role of rare cells and behaviors within the complex systems, and the dynamic nature of the response networks. *Descriptive* single cell sequencing cannot, in isolation, identify the dynamic cellular functions that drive tissue development and disease. In order to move from *descriptive* studies towards *explanatory and predictive* models, we need a better link between systems level molecular phenotypes and the underlying dynamic microenvironment-driven functions. A systems-level approach is necessary to identify nodes and times within the interconnected network of interactions that drive and control emergent tissue behaviors (**Figure 1**).

Here, we provide a perspective on utilizing recent technological advances in the fields of single cell systems biology, genetic engineering, and high content imaging to link dynamic single cell microenvironment-driven functions to their molecular drivers within engineered microenvironments. As biomaterial and organoid microsystem technologies advance, they increasingly bridge the complexity gap between reductionist 2D in vitro cell culture and complex in vivo systems (**Figure 1**). This intermediate complexity allows for the *in vitro* production of realistic models, while allowing for control and interrogation in ways that would not be possible *in vivo*. We detail a systems tissue engineering framework for exploring sources of complexity within engineered microsystems including i) the dynamics of tissue development and dysregulation, ii) emergent heterogeneity and the function of rare cell types through single cell analysis, and iii) the spatial distribution of individual cells controlling structure and function within a tissue.

**2. Capturing dynamics within engineered microenvironments**

Tissue behaviors emerge from many layers of dynamic cellular processes including cell cycling, migration, and matrix remodeling with time scales ranging from seconds to days, each regulated by subcellular molecular processes including protein-protein interactions and gene expression with timescales from microseconds to hours (18). Epigenetic changes from these inputs may occur over days or longer guiding long-term cellular processes like cellular differentiation(19, 20). The dynamic processes are ultimately integrated to regulate cell and tissue behavior. Recording the sequence of molecular regulatory events and the corresponding tissue properties would facilitate the identification of the key pathways that are driving tissue function across the stages of tissue development or disease progression. Biomaterial microsystems provide a means of creating systems that can be dynamically imaged to report on cell differentiation and maturation, and ultimately tissue function. Here we outline emerging methods and opportunities for dissecting dynamic tissue behaviors using biomaterial microsystems.

*2.1 Dynamic microenvironments*

Engineered 3D hydrogels and microsystems with the ability to analyze dynamics would facilitate the identification of the key drivers of tissue function. In transitioning cells from 2D tissue culture plastic to 3D culture, particularly within systems that can be remodeled, cells experience dynamic changes in geometry, structure, and composition of their environment. While the biochemical environment can be changed on demand without disrupting adherent cells by changing culture media, more recently, hydrogels have been designed whose biomechanical or biochemical properties can be switched on demand via an external signal, such as light (21, 22). Cells across multiple material platforms respond with dramatic changes in behavior after altering the mechanical stiffness of the underlying substrate (23-26). These systems for modulating the signals on demand are enabling for analysis of dynamic cellular responses. More recent biomaterial designs are supporting fully reversible(27) mechanical or biochemical changes (28), allowing for *in situ* study of the dynamics of microenvironmental memory (29, 30) and its role on cell and tissue function. Furthermore, materials may sense and respond to their microenvironment, thereby providing biomimetic feedback loops to resident cells (31, 32). These material systems provide critical tools for probing how dynamic microenvironmental changes control systems-level tissue function.

*2.2 Dynamic cell and tissue monitoring with genetic reporters*

The accessibility of biomaterial microsystem constructs to imaging is a key feature for analyzing dynamic responses. Optically clear biomaterial culture systems (33, 34) with homogenous microstructure (yet potentially with engineered nanostructures with length-scales less than half the wavelength of the light used for imaging) can transmit light much more efficiently than those containing high refractive index microscale fibers (35, 36) and micron-scale features which scatter light (37). Light microscopy has long been employed for identifying cell and tissue structures and their dynamics, the spatial position of cells relative to each other, or intracellular distribution of organelles and proteins. Connecting cell and tissue structure and function to dynamic changes in molecular state is uniquely achieved by live cell imaging of genetic reporters.

With the use of genetic reporters, fluorescence and luminescence can be employed to capture systems-level dynamics in molecular signaling, with emerging systems enabling large scale analysis. Analytical assays such as RNAseq and proteomics require destructive snapshot measurements, and while these techniques can be performed over a time-course experiment, they must be performed on replicate experiments of unique and distinct sets of cells and in themselves do not allow for direct correlation with tissue function. These limitations can be especially problematic when working with high complexity 3D biomaterial microsystems. Live cell imaging of genetically-encoded reporters can uniquely monitor the dynamic signaling within living cells or tissues. High quality reporters have been generated for a variety of intracellular processes, including transcription factors (38-47), microRNA (48-51), kinase activity (52-54), metabolism (55, 56), chromatin organization (57), and protein-protein interactions (58, 59) among others. Scaling dynamic genetic reporters to dozens of reporters in a single experiment has been achieved by pre-manufacturing libraries of ready to use reporter transduction agents, such as lentivirus, that can be used to transduce cells in a parallel format (40, 44-46, 51, 60, 61). The transduced cells are then cultured and imaged dynamically in high-content array to investigate the dynamic molecular response networks triggered by micro-environmental stimuli. Using microwell plates with each well containing a distinct reporter, the signaling induced by RGD ligand or mechanical stiffness was investigated through identifying the transcriptional response networks of 50 transcription factors. These studies identified transcription factors that are specific to ligand density or mechanical stiffness, as well as transcription factors common to both stimuli (45). Here, bioluminescence imaging of each well reported population level responses, yet was unable to capture cell specific responses.

Recent technological advances have furthered the potency and scalability of genetic reporters for monitoring dynamic cellular processes. Luminescent proteins have been attractive for detecting low levels of transcription factor activity, because the low background imparts a high signal to noise ratio (62, 63). Advances in luminescent reporter proteins, including improvements in brightness (64) and in red-shifting for better tissue penetration and color multiplexing (65-68), allow them to outperform fluorescent proteins for reporting on whole-tissues or multicellular systems. Pairing the low read noise of Electron Multiplying Charge-Coupled Device (EMCCD) or scientific Complementary Metal-Oxide-Semiconductor (sCMOS) cameras with an appropriate objective and tube lens can enables bioluminescence microscopy (69, 70). Nonetheless, fluorescent proteins with their much greater brightness still have advantages for single cell and 3D reporter imaging. Advances in commercial DNA synthesis (71, 72) allow for the straightforward and rapid synthesis of the genetic parts, with components assembled into multi-kilobase sequences with robust methods like Gibson Assembly (73). Further, advances in genetic engineering including the advent of CRISPR cas9 technology (74, 75) and engineering of cas9 variants with improved specificity (76, 77) allows for scalable strategies for reporting on activity of endogenous promoters and enhancers (78-82), or for the insertion of large multi-kilobase (83) genetic cassettes capable of bearing several reporters at a genetic safe harbor site.

## 3. Capturing the role of single cell heterogeneity in emergent tissue behaviors

Heterogeneity of individual cells within a tissue is known to be of great clinical significance, as illustrated by intratumoral heterogeneity imparting resistance to cancer therapeutics (84). On a more fundamental level, heterogeneity may facilitate basic tissue functions including fate plasticity and information coding (85). Dynamic responses that appear continuous at the population scale can be heterogeneous and switch-like at the single cell level (86), indicating there is much to learn about how single cells make functional decision. However, linking systems scale single cell molecular information to the single cell functions within complex biomaterials microsystems requires experimental information linking molecular form to single cell function. New technologies including single cell RNA sequencing (87-89) provide molecular phenotype, and have revolutionized our understanding of the complexity and distribution of single cell phenotypes within *in vivo* tissues. Here, we describe approaches that aim to bridge single cell genomics technologies with high content imaging to capture the contribution of cellular heterogeneity or the function of rare cell populations on collective microtissue function (**Figure 2**).

*3.1 Engineered biomaterial microsystems for single cell analysis*

Biomaterial platforms provide a means by which to elicit and observe single cell functional responses to controlled environments. For example, mature platforms for applying concentration gradients of soluble factors to cells in 3D culture allow observation of heterogeneous single cell chemotaxis responses of individual cells (90-93), with some rare cells displaying exceptional chemotaxis. Likewise, manipulation of the mechanical properties and structure of the ECM have allowed for dissection of individual cells exerting forces and migrating in response to the ECM (35, 36, 94-98). Single cell force generation can vary by orders of magnitude between cells and microenvironments (36, 99-101). Using microsystems, the relative role of soluble and cell-cell contact factors in driving behavior has been explored in different cell types including single NK cell activation (102) and killing of cancer cells (103). Single NK cell cytotoxicity experiments have demonstrated rare NK cells are responsible for the majority of cytotoxic activity against cancer cells (104-108). Across these systems, a vast heterogeneity of single cell behaviors in both time and space has been observed, which points to the importance of single-cell analysis as well as the essential role of dynamic cell behaviors in the overall structure and function of tissues.

*3.2 Single cell pseudo-temporal omics analysis*

Single cell transcriptomic and multi-omic technologies provide a means to capture a detailed molecular *description* of cell heterogeneity at any given time within engineered microenvironments. Single cell RNA sequencing technology (87-89) has scaled exponentially over the last decade (109) and now has a well-established ecosystem of analysis tools (110-112). Further, multi-omic technologies provide a means to analyze non-transcriptionally controlled pathways by integrating single cell transcriptomic data with single cell proteomic and chromatin accessibility measurements (113). CITE-seq (114) and REAP-seq (115) can simultaneously quantify mRNA and extracellular protein content in individual cells by using antibody cocktails barcoded with oligonucleotides which are sequenced along with endogenous mRNA. A complementary tool is the CRISPR loss of function screens, where a library of many guide RNA's can be used to knock down or knock out genes in individual cells, allowing for detailed study of the role of proteins in gene regulatory networks(116-119). Other sequencing modalities including single cell ATAC-seq (120, 121) and THS-seq (122) can be used to link single cell chromatin accessibility to the state of transcriptome of individual cells. Analysis methods for these genomics datasets provide *descriptive* clustering of cells into categories by their gene expression, but they cannot directly link gene expression to cell behavior without additional information. Information including single cell spatial location and orientation, single cell functional phenotype, and dynamic changes of individual cells over time are lost when the tissue is processed for sequencing.

Moving toward a *predictive* understanding of cellular heterogeneity and tissue function will require a means of linking and analyzing snapshot *descriptions* of single cell molecular phenotypes across time. Pseudo-temporal analysis of scRNAseq datasets(113), with methods such as pioneering work Monocle (111), achieve this goal by creating single cell trajectories across dynamic processes like differentiation (111, 112, 123-125) and the cell cycle(124). Here, cells are harvested at several time points during a dynamic process such as cellular differentiation from replicate experiments, and scRNAseq or another omics method is conducted on the cells harvested at each timepoint. The transcriptome of each individual cell across replicate experiments harvested at distinct time points is then ordered on a trajectory of the biological process occurring over time, with this ordering referred to as psuedotime. The ordering and branchpoints of graph networks produced across pseudotime are then used to predict critical molecular regulators and decision points. While original algorithms allowed for only linear trajectories, newer methods support cycling and circular paths (113). These methods are well suited for the study of dynamic heterogenous cellular responses in the microenvironment of biomaterial microsystems. While dynamic information can be captured by pseudo-temporal ordering, this ordering is obtained from the unique transcriptome of individual cells obtained from replicate experiments sequenced at multiple times; each individual cell is measured only once, and essential dynamic and spatial information is inherently lost in this process.

*3.3 Connecting tissue function to molecular drivers with live single cell imaging*

Directly linking dynamic single cell function to dynamic single cell molecular phenotypes requires observation of individual live cells over time. Modern high content live cell imaging assays can track morphology and biomarkers of many thousands of individual cells across time as they respond to stimuli from their microenvironment (126, 127). In order to provide information on molecular phenotype, cell lines or progenitor cells (128-130) can be produced containing genetic reporter elements using fluorescent or luminescent proteins. An automated live cell microscope equipped with an incubated stage can image individual cells across many wells of a microtiter plate over time, recording changes in single cell behavior simultaneously with changes in single cell reporter activity. Image analysis codes are then applied to link, correlate, and interrogate the order of changes in dynamic single cell functions with dynamic single cell molecular phenotype. For example, high content imaging of genetic reporters has been employed to screen the effects of thousands of drugs on cell function (131), and to elucidate the single cell toxicity pathways activated by different drugs (132). Here, genetic reporters can be included in one or more cellular subsets of the microenvironment to facilitate imaging and record their specific contribution to the emergent behavior. These methodologies are directly applicable to the study of cells on planar 2D interfaces within biomaterials where multiple cell types or variation in the local microenvironment such as material or soluble gradients are present. We note that the dynamic physical movements and behaviors of individual cells residing within biomaterial microsystems are also routinely recorded to connect material microenvironment to cell function. Integration dynamic imaging of genetic reporters within such systems completes the linkage from microenvironment to function to molecular drivers within individual cells.

A staged discovery strategy may be necessary for incorporating single cell genomics and high content imaging of genetic reporters to efficiently link single cell functions to their molecular phenotypes. Here, an initial round of single cell genomics and pseudo-temporal analysis would be used to identify regulators stratifying the dataset. Then high content imaging on a more focused set of live cell genetic reporters can connect dynamic microenvironment driven functions to the larger systems-level molecular phenotypes. These methods can be computationally integrated with targets, with subsequent validation, such as through the use of high-throughput loss of function screening.

3.*4 High content single cell imaging for 3D systems*

Most cells live within a fully 3D rather than a 2D planar geometric context, and 3D cell culture is necessary for obtaining some complex phenotypic responses (133, 134). However, 3D systems present unique challenges for high content imaging experiments including the requirement to acquire and process large 3D datasets, noise and bias from imaging through hydrogel or ECM material, and the requirements for highly specialized imaging systems for optimal performance. Conventional imaging systems including wide-field and confocal point scanning have substantial limitations for dynamic imaging of 3D microenvironments including phototoxicity and photobleaching as they illuminate through the whole sample (135). Light sheet microscopy systems provide the unique ability to illuminate only the current imaging plane, greatly reducing phototoxicity in 3D imaging studies. Recent refinements including the advent of lattice light sheet microscopy (136) and the subsequent integration of aberration-correction adaptive optics (137) may further enhance performance. However, conventional light sheet microscopy requires complicated mounting schemes and small samples owing to the requirements of 2 orthogonal light paths. Recent work toward open top light sheet microscopy (138-140) and oblique plane microscopy (141) would allow for plate-based multi-well imaging enabling truly high content 3D imaging studies. For more on the current state of high-content cell imaging we refer the reader to an excellent review (126).

## 4. Capturing spatial information driving tissue behavior

The spatial distribution of cells and features within and between niches in a tissue are integral for controlling collective tissue behavior and function (142-144). Coordinated behaviors within tissues may at least in part ultimately emerge from changes in gene expression within subsets of spatially defined cells (145). Cells communicate and provide spatial signals to neighbors through concentration gradients (146) and waves (147), cell-cell contacts, ECM remodeling, and mechanical forces (148, 149). Spatial patterning produced by biomaterial microsystems can be used to recreate complex tissue architectures that would not be otherwise possible (150). For example, geometric micropatterning of cell adhesion (151, 152) can drive a complex multi-cellular emergent behavior, such as where cancer cell stemness is concentrated at the geometric edges and vertices of colonies (153). Patterning of gene delivery has been employed to control concentration profiles that spatially oriented neurite extension (154, 155). Similarly, microfluidic generated concentration gradients (156-158) and material stiffness gradients (99, 159, 160) induce chemotaxis and durotaxis respectively in diverse cell types. Patterning of mechanical interfaces guide development of microtissues including the spontaneous polarized development of amnion-like tissue from stem cells (161). We expect the use of biomaterial microsystems for control of spatial patterning of the tissue microenvironments paired with emerging spatially resolved sequencing (162-164) will further resolve the mechanisms and role of spatial signaling networks in tissue behaviors.

*4.1 Spatially resolved omics within engineered microsystems*

The most conceptually straightforward approach to link single cell microenvironment-driven function to a systems-level molecular phenotype is to isolate and index individual cells from culture after observing their behavior and function. Single cell isolation (165, 166) can be achieved by physically picking cells using a micromanipulation system, with microfluidic systems (167), or by culturing cells in isolation in microwells. However, current systems for cell picking by micromanipulation remain low throughput and labor intensive to operate (155, 165, 166). This approach can however be effective for experiments with low cell numbers, as cells can be directly picked from complex multicellular microenvironments. Isolated culture approaches in contrast can be scaled to 10's or 1,000's of individual cells using microfluidic technology(167), but an inherent limitation is that cells will not receive the cell-cell contacts and soluble factors from other cells in their microenvironment. In one isolation culture approach, researchers used an ECM coated Fluidigm C1 chip to isolate individual cells into microwells then applied fluorescence microscopy to track NF-kB transcription factor reporter activity over time after lipopolysaccharide stimulation(168). The chip was then used to create single cell cDNA libraries for single cell RNA sequencing providing both a history of dynamic NF-kB activity and transcriptome data for each individual cell. This capacity to link a history of dynamic functional behavior to the full transcriptome within the same individual cell is unique to the physical separation approach and allows for a direct link between microenvironment driven function and molecular phenotype.

Emerging methods to link spatial position and gene expression include MERFISH(162), seqFISH(163), and STARmap(164), which can detect from 100's (162, 163) to up to 1000(164) genes per individual cell *in situ* within fixed tissue samples. This *in situ* approach would allow for dynamic monitoring of the sample before fixation to connect past behaviors of each cell to the molecular phenotype and spatial information in the sample at the time of fixation. However, current methods remain complex, time consuming, and low throughput, requiring dedicated facilities and expertise (169-171). In STARmap, DNA nanoballs are produced locally at the site of each individual cell and entrapped within a 3D DNA hydrogel. Advances in nanoparticle and hydrogel technologies will be expected to facilitate improvements in the in situ sequencing technology. Even with all of the spatial information preserved, in situ spatial genomics remains a destructive snapshot technique and cannot give information about dynamic cell signaling networks.

*4.2 Biomaterials and microsystems for spatial interrogation and control*

Continuing advances in biomaterial microsystems uniquely allow for spatial control of cells to interrogate the role of spatial organization on tissue function. Synthetic hydrogel chemistries for 3D cell culture allow for high-resolution patterning of degradation (23), mechanical stiffness/crosslinking (172), and protein ligands(173) using photolithography methods. Just as materials can be patterned with light, optogenetic technology, where light and photosensitive proteins are used to control gene expression (174, 175), can be used to photo-pattern gene expression within inhabitant cells of a microtissue (176). Advances in 3D printing biomaterial microsystems allow for the patterning and spatial organization of several biological inks containing multiple cells and/or materials and voids in 3D (177-179). Patterning systems have also recently extended into the temporal domain with one system describing a method to simultaneously pattern 3 proteins, with spatial and temporal control, within a biomaterial (180). We anticipate that continuing advances (33, 181) in the spatiotemporal control of biomaterial microsystems will enable mechanistic studies of the spatial coordination and feedback loops driving tissue behaviors.

## 5. Conclusion

The integration of biomaterial technologies with genomics and high content imaging offers the opportunity to molecularly dissect normal and abnormal tissue responses, and ultimately to develop the design principles for biomaterial systems that direct cellular processes and tissue formation. This combination of techniques offers the opportunity to describe the complex spatio-temporal processes at the cellular scale, while also capturing the heterogeneity of cellular responses, with applications in basic science, drug development, and tissue engineering (150, 182, 183). Given the complexity of emergent tissue behaviors, the path toward designing improved biomaterial microsystems will likely involve the application of systems-level experimental workflows that can elucidate the role of dynamics, spatial coordination, and rare cells in driving tissue functions (**Figure 2**). This overview of systems level analysis and the currently available tools highlights opportunities for innovation in the biomaterial platforms and their integration with computational algorithms that can process the data and identify the design parameters, which will advance the understanding of biomaterial function and potentially enhance the ultimate pace of translation.

**FIGURES**

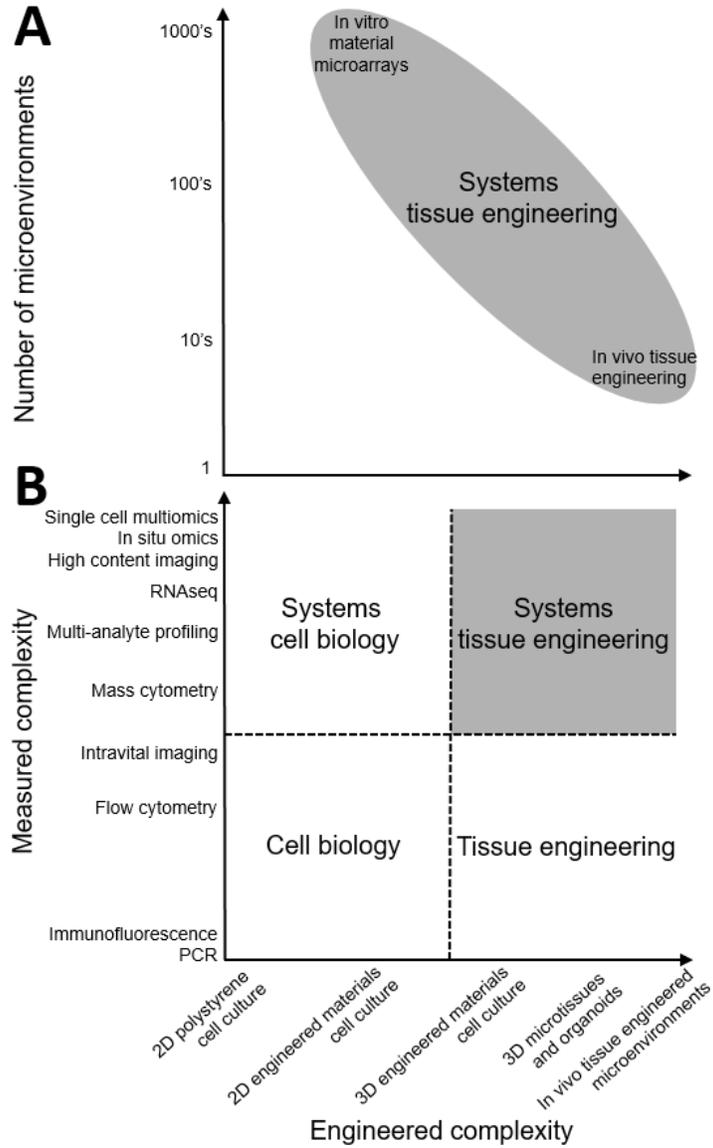

**Figure 1. Systems tissue engineering for study of complex microenvironments.**

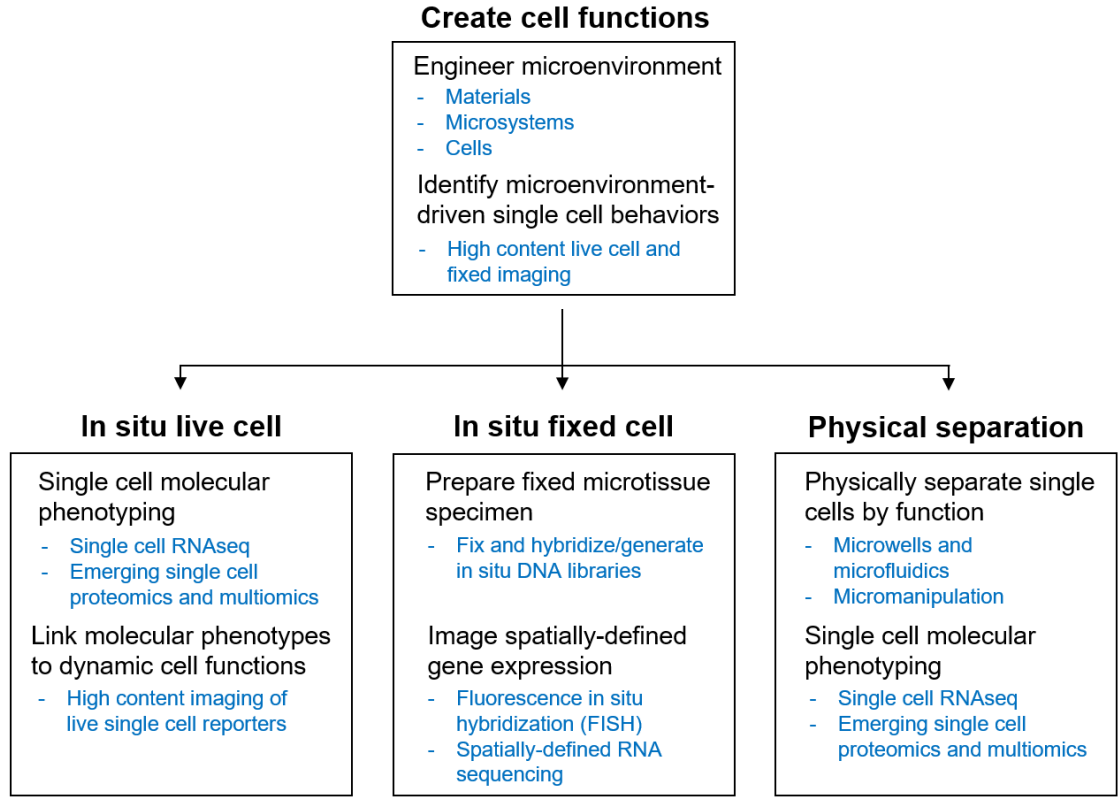

**Figure 2: Strategies for linking single cell functions to their molecular phenotype within engineered microenvironments.**

130. Bressan RB, *et al.* (2017) Efficient CRISPR/Cas9-assisted gene targeting enables rapid and precise genetic manipulation of mammalian neural stem cells. *Development* 144(4):635.
131. Kang J, *et al.* (2015) Improving drug discovery with high-content phenotypic screens by systematic selection of reporter cell lines. *Nature Biotechnology* 34:70.
132. Wink S, Hiemstra S, Herpers B, & van de Water B (2017) High-content imaging-based BAC-GFP toxicity pathway reporters to assess chemical adversity liabilities. *Archives of Toxicology* 91(3):1367-1383.
133. Griffith LG & Swartz MA (2006) Capturing complex 3D tissue physiology in vitro. *Nat Rev Mol Cell Biol* 7(3):211-224.
134. Edmondson R, Broglie JJ, Adcock AF, & Yang L (2014) Three-dimensional cell culture systems and their applications in drug discovery and cell-based biosensors. *Assay Drug Dev Technol* 12(4):207-218.
135. Frigault MM, Lacoste J, Swift JL, & Brown CM (2009) Live-cell microscopy – tips and tools. *Journal of Cell Science* 122(6):753.
136. Chen B-C, *et al.* (2014) Lattice light-sheet microscopy: Imaging molecules to embryos at high spatiotemporal resolution. *Science* 346(6208).
137. Liu T-L, *et al.* (2018) Observing the cell in its native state: Imaging subcellular dynamics in multicellular organisms. *Science* 360(6386).
138. McGorty R, *et al.* (2015) Open-top selective plane illumination microscope for conventionally mounted specimens. *Optics express* 23(12):16142-16153.
139. McGorty R, Xie D, & Huang B (2017) High-NA open-top selective-plane illumination microscopy for biological imaging. *Optics express* 25(15):17798-17810.
140. Glaser AK, *et al.* (2017) Light-sheet microscopy for slide-free non-destructive pathology of large clinical specimens. *Nature Biomedical Engineering* 1:0084.
141. Dunsby C (2008) Optically sectioned imaging by oblique plane microscopy. *Optics express* 16(25):20306-20316.
142. O'Brien LE & Bilder D (2013) Beyond the niche: tissue-level coordination of stem cell dynamics. *Annu Rev Cell Dev Biol* 29:107-136.
143. Park S, *et al.* (2017) Tissue-scale coordination of cellular behaviour promotes epidermal wound repair in live mice. *Nature cell biology* 19(2):155-163.
144. Kim EJY, Korotkevich E, & Hiiragi T (2018) Coordination of Cell Polarity, Mechanics and Fate in Tissue Self-organization. *Trends in Cell Biology* 28(7):541-550.
145. Featherstone K, *et al.* (2016) Spatially coordinated dynamic gene transcription in living pituitary tissue. *eLife* 5:e08494-e08494.
146. Carmona-Fontaine C, *et al.* (2017) Metabolic origins of spatial organization in the tumor microenvironment. *Proc Natl Acad Sci U S A* 114(11):2934-2939.
147. Deneke VE & Di Talia S (2018) Chemical waves in cell and developmental biology. *The Journal of Cell Biology* 217(4):1193.
148. Reinhart-King CA, Dembo M, & Hammer DA (2008) Cell-Cell Mechanical Communication through Compliant Substrates. *Biophysical Journal* 95(12):6044-6051.
149. Winer JP, Oake S, & Janmey PA (2009) Non-Linear Elasticity of Extracellular Matrices Enables Contractile Cells to Communicate Local Position and Orientation. *Plos One* 4(7).
150. Dye BR, *et al.* (2016) A bioengineered niche promotes in vivo engraftment and maturation of pluripotent stem cell derived human lung organoids. *eLife* 5:e19732.
151. Chen CS, Mrksich M, Huang S, Whitesides GM, & Ingber DE (1997) Geometric control of cell life and death. *Science* 276(5317):1425-1428.
152. Théry M (2010) Micropatterning as a tool to decipher cell morphogenesis and functions. *Journal of Cell Science* 123(24):4201.
153. Lee J, Abdeen AA, Wycislo KL, Fan TM, & Kilian KA (2016) Interfacial geometry dictates cancer cell tumorigenicity. *Nat Mater* 15(8):856-862.
154. Houchin-Ray T, Whittlesey KJ, & Shea LD (2007) Spatially Patterned Gene Delivery for Localized Neuron Survival and Neurite Extension. *Molecular Therapy* 15(4):705-712.
155. Houchin-Ray T, Swift LA, Jang J-H, & Shea LD (2007) Patterned PLG substrates for localized DNA delivery and directed neurite extension. *Biomaterials* 28(16):2603-2611.
156. Kim BJ & Wu M (2012) Microfluidics for Mammalian Cell Chemotaxis. *Annals of biomedical engineering* 40(6):1316-1327.